\documentclass[showpacs,twocolumn,prl,amsmath,amssymb]{revtex4}
\usepackage{graphicx}
\usepackage{amsmath}
\usepackage{amssymb}

\begin{document}
\title{Globally coupled chaotic maps and demographic stochasticity}
 \author{David A. Kessler}
 \email{kessler@dave.ph.biu.ac.il}
\affiliation{Department of Physics, Bar-Ilan University,
 Ramat-Gan 52900 Israel }
\author{Nadav M. Shnerb}
\email{shnerbn@mail.biu.ac.il}
\affiliation{Department of Physics, Bar-Ilan University,
 Ramat-Gan 52900 Israel }

\begin{abstract}
The affect of demographic stochasticity of a system of  globally
coupled chaotic maps is considered. A two-step model is studied,
where the intra-patch chaotic dynamics is followed by a migration
step that coupled all patches; the equilibrium number of agents  on
each site, $N$, controls the strength of the discreteness-induced
fluctuations. For small $N$ (large fluctuations) a period-doubling
cascade appears as the coupling (migration) increases. As $N$ grows
an extremely slow dynamic  emerges,  leading to a flow along a
one-dimensional family of almost period 2 solutions. This manifold
become a true solutions in the deterministic limit. The degeneracy
between different attractors that characterizes the clustering phase
of the deterministic system is thus the $N \to \infty$  limit of the
slow dynamics manifold.
\end{abstract}


\maketitle

The dynamics of coupled chaotic maps have attracted a lot of
interest in the last decades, following the pioneering works of
Kaneko~\cite{kaneko, kaneko1}. A substantial part of the study is
focused around the paradigmatic model of globally coupled maps,
where many fundamental results like mutual synchronization,
dynamical clustering and glassy behavior were
demonstrated~\cite{mzm}.  The universal character of the chaotic
dynamics makes the coupled maps model relevant to many phenomena,
ranging from neural systems and human body rhythms  to coupled
lasers and cryptography \cite{syn}.

Here we consider the effect of demographic stochasticity (shot
noise) on various phases of a globally coupled system. This problem
emerges naturally while applying the theory to spatially extended
ecologies.

Many old \cite{gause} and recent  \cite{kerr} experiments suggest
that the well-mixed dynamics of simple ecosystems (single species or
victim-exploiter system) are extinction-prone, and that the system
acquires  stability only due to its spatial structure, a result
supported also by numerical simulations of many models
\cite{durrett, x1}. The extended (spatial) system survives due
to the possibility of migration among spatial patches. This
migration should be large enough to allow for recolonization of
empty patches be emigrants. On the other hand \cite{earn}, too much
migration is also dangerous, as it leads to global synchronization,
in which case the system acts essentially as a single, well-mixed
patch, with its vulnerability to extinction.

The globally coupled system, which obeys,
\begin{equation} \label{1}
s^i_{t+1} = (1-\nu) F(s^i_t) + \frac{\nu}{L} \sum_{j \neq i}
F(s^j_t),
\end{equation}
 is a natural and popular framework to discuss the dynamics of so-called meta-populations \cite{hanski} on various
 patches with migration between the patches~\cite{earn}.
 Here,   $s_i$ is the population density on the
$i$-th site and $F$ is the chaotic map. $\nu$ is the migration
parameter (the chance of an individual agent to leave its site) and
$1\le i \le L$, where $L$ is the number of patches. To address the
problem of extinction properly, however, it is necessary to account
for the discrete nature of the population and the absorbing
character of the zero population state. This is achieved by the
addition of demographic (shot) noise to the system. Clearly, if the
local populations are all large, this effect is tiny. However, for
small and moderate populations, the effects as we shall see can be
very important and give rise to new phenomena.  This is somewhat
surprising, since it is widely assumed that chaos generates its own
noise, and the addition of other noise should not induce a
qualitative change in the dynamics. However, one important feature
of coupled maps is the appearance of attractive regular orbits for
certain value of the coupling; it turns out that once the system is
kicked from these orbits it follows a long excursion on its way back
(this issue will be discussed in a separate publication).  The
presence of continuous noise may thus change drastically the
observed dynamics. The results of the deterministic theory (like
clustering) should reappear, though, in the large $N$ limit, where
$N$ is the number of particles at a patch.

Let us begin our discussion by specifying a  stochastic model. Again
we have a collection of $L$ sites, where the local population density $s_i^t$
is now replaced by an integer  $n_i^t$. The update proceeds in two
steps. First, the reproduction and competition generates a new value
of $n_i$. This value is taken to be drawn from a Poisson
distribution with mean $F(n_i)$, where $F(n)$ is a chaotic map.  In
this paper, we take as our choice of $F$ the paradigmatic Ricker map
\cite{ricker}, $F(n)=rne^{-n/N}$, where $r$ is the growth factor.
Also, we have fixed the value of $r=20$, which is well in the
chaotic regime of the deterministic map. (We have  chosen the Ricker
dynamics just because it simplifies our numerics;  our results hold
for a broad range of different maps, and we believe that any chaotic
map, including the logistic one, should exhibit similar behavior.)
The second phase is the dispersal phase, in which with probability
$\nu$, each of the inhabitants of every site can decide to leave and
pick a new site at random.  This model is essentially similar to
that used by Hamilton and May \cite{hm} to study optimal dispersal
rates, except for the chaotic nature of the on-site
reproduction/competition dynamics of the present model.

We have simulated this system directly using Monte-Carlo technique
for $L=10000$. However for the purpose of analysis, it is more
convenient to study, as do Hamilton and May, the $L\to \infty$
limit.  This limit is completely characterized by a probability
distribution $\psi_n^t$, the chance of a given site to have $n$
individuals at time $t$.  The probability of having $m$ individuals after birth and competition
is fixed by $n$, and the probability distribution for the number of individuals leaving to other sites is then fixed by $m$.
Since we are dealing with an infinite reservoir, the probability distribution for the number of incoming individuals is
fixed by $\lambda$, the average population after the birth/competition phase, which in turn is fixed by $\psi$:
\begin{equation} \label{3}
\lambda = \overline{F(m)} = \sum_{m=0}^\infty \psi_m F(m).
\end{equation}
Accordingly, $\psi$ dynamics follows,
\begin{equation} \label{2}
\psi_n^{t+1} = \sum_{m=0}^\infty \psi_m^t e^{-\mu(m)}
\frac{\mu(m)^n}{n!} = {\cal{M}}_{nm}\psi_m^t
\end{equation}
where $\mu(m) = F(m)(1-\nu) + \nu \lambda$.
The update rule is a linear transformation of the probability vector
$\psi$ which conserves probability.  The transformation matrix
$\cal{M}=\cal{M}(\lambda)$ is a functional of the input state, since it depends on
$\lambda$, which depends of $\psi$.  It is this
dependence on the input state that renders the problem nonlinear,
and gives it its rich structure.

We consider first a constant (period 1) solution. Picking an
arbitrary value for $\lambda$ one gets a Markov matrix
$\cal{M}(\lambda)$ from (\ref{2}). Due to the Markov property this
matrix must admit an invariant eigenvector (a right eigenvector with
eigenvalue 1) $\psi_n(\lambda)$. This procedure is consistent
\textit{iff} $\psi_n(\lambda)$ and $\lambda$ satisfy  Eq. (\ref{3}).  This consistency condition
determines $\lambda$ for the period 1 solution, which in turn determines $\psi$ of the
period 1 solution.
A period 2 solution is obtained in a similar way: Picking arbitrary
values for $\lambda_1$ and $\lambda_2$ the Markov matrix $
{\cal{M}}(\lambda_2){\cal{M}}(\lambda_1)$ must admit an invariant
eigenvector $\psi_n$. The solution is consistent \textit{iff} the system
satisfies  the two auxiliary conditions $\lambda_1 = \sum \psi_n
F(n)$ and $\lambda_2 = \sum  \left[{\cal{M}}(\lambda_1) \psi
\right]_n \ F(n)$. Extending this procedure one may find orbits of
higher periodicity by searching through the space of quartets and
octets of $\lambda$-s with the appropriate auxiliary conditions.

\begin{figure}
\vspace*{-.4in}\includegraphics[width=0.47\textwidth]{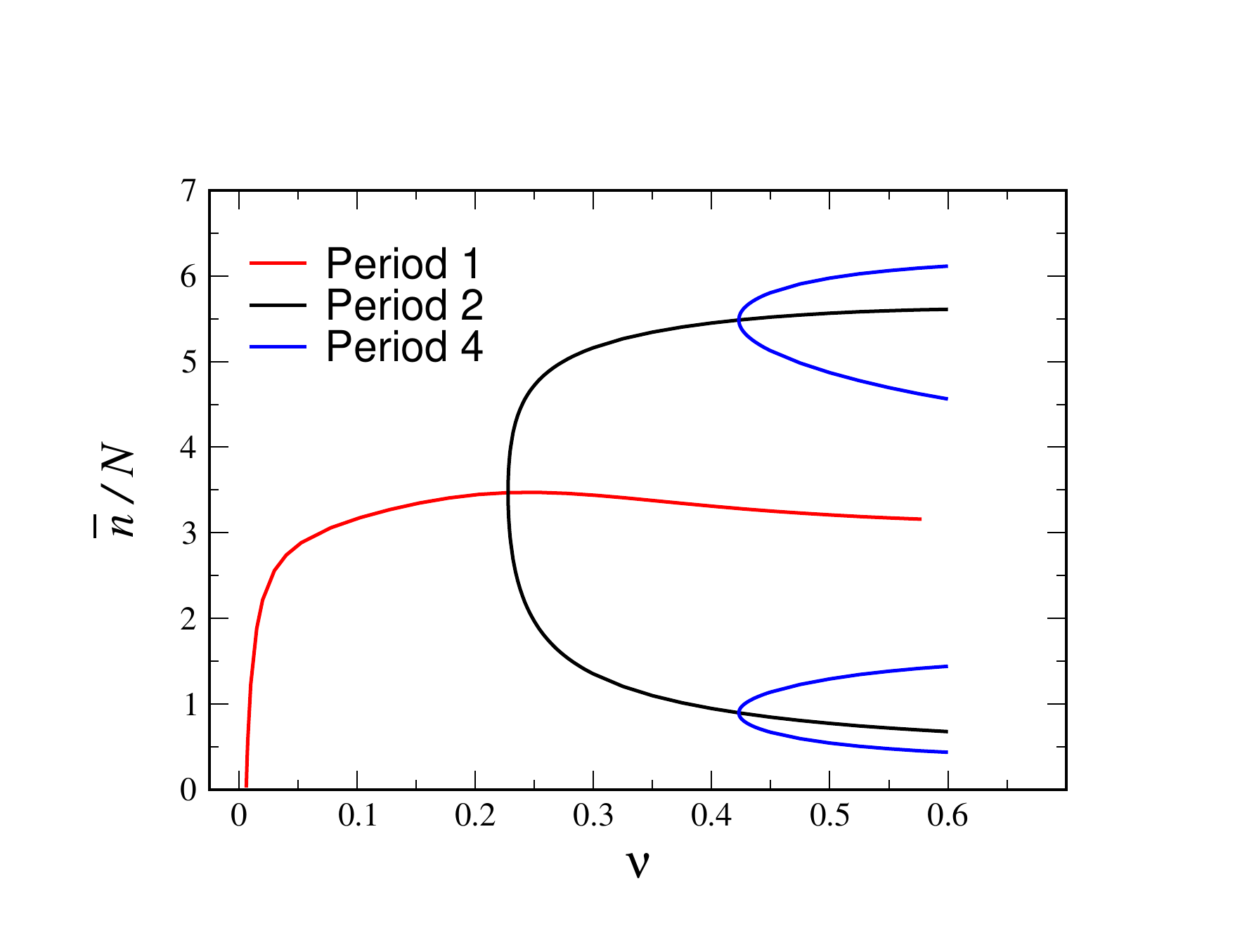}\\
\vspace*{-.5in}\includegraphics[width=0.47\textwidth]{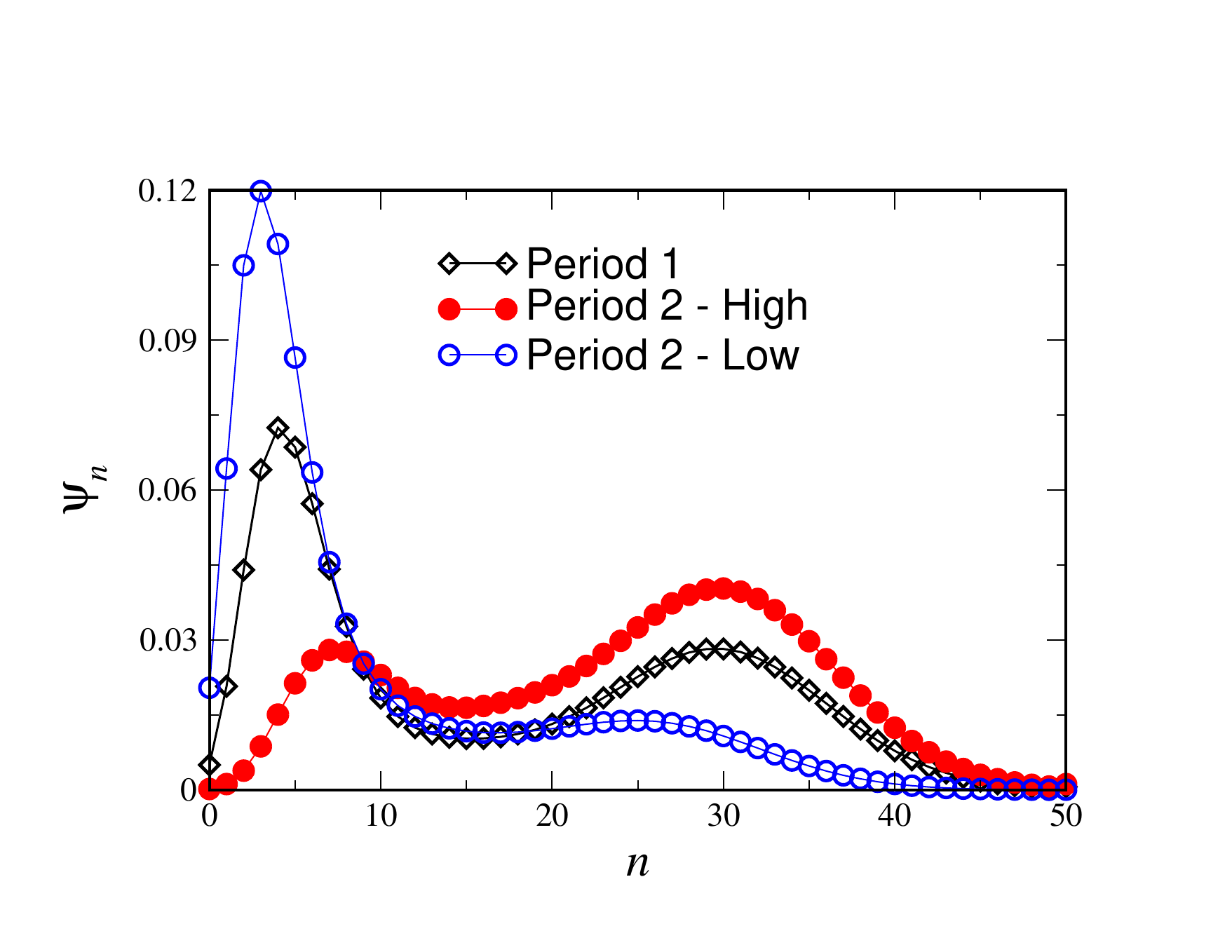}\vspace*{-.2in}
\caption{Upper:  Solution branches for $N=5$, showing the period 1,
2 and 4 branches as a function of $\nu$.  Lower: Probability
Distributions $\psi_n$ for the period 1 solution at $\nu=0.2$ and
for the two phases of the period 2 solution at $\nu=0.25$. The
points are connected to guide the eye.} \label{fig1}
\end{figure}

One trivial solution always exists for period 1 orbits: the
absorbing state for which $\lambda = 0$ and $ \psi_n = \delta_{n0}$.
It turns out that for small values of $\nu$ this is the only
solution and  the probability vector converges exponentially quickly
to $\delta_{n0}$ (see Figure \ref{fig1}, where the results for
strong stochasticity, $N=5$, are summarized). This is as expected,
since there is a finite probability of an individual site to go
extinct and without sufficient dispersal to enable recolonization,
more and more sites go extinct as time goes on. Above a critical
value of $\nu$ the system quickly adopts a nontrivial period-1
configuration. The function $\overline{n}(\nu)$ is plotted in Figure
\ref{fig1}. The corresponding distribution $\psi_n$  is
characterized by two peaks, as exemplified in the lower panel of
\ref{fig1}. It shows two peaks, each of which has half the
probability. The system decomposes into two clusters that oscillates
$180^\circ$  out of phase with respect to the other. Since the two
clusters have equal weight, the overall occupancy of the system is
time-independent.

What about the stability of the period-1 solution? The above numerical
technique  of course only identifies solutions and does not say anything about their stability. One can prove,
however, that any periodic orbit must become unstable as $\nu \to 1$.
As pointed out by Durrett and Levin \cite{durrett}, in that case the
occupation of a site just before the reaction step is  a
Poisson distribution with a mean given by the total population in the last
step. As a result, $\overline{n}$ satisfies the  iterative map,
\begin{equation}
\overline{n}_{t+1} = \sum_{k} k F(k) e^{-\overline{n}_t} \frac{(\overline{n}_t)^k}{k!}\end{equation}
where $F$ is the deterministic map. In the Ricker case the resulting
map for $\overline{n}$ is also unimodal and the resulting transformation
is chaotic in the regime of parameters considered here.  Thus any periodic orbit must lose its stability as $\nu$ approaches unity.

All the considerations and the results presented so far are general
and are independent of the strength of the stochasticity, which is
inversely proportional to $N$.  We turn now to consider the
differences between strong and weak stochasticity and the
semi-deterministic (large $N$) limit.

Let us refer again to Figure \ref{fig1} where the results for $N=5$
(strong stochasticity) are graphed. There is a range of $\nu$ for
which the period 1 is stable, as can be verified by MC simulation.
Increasing $\nu$  this time-independent state goes unstable and the
system undergoes a forward (supercritical) bifurcation to a
period-two state. This state losses stability in favor of a period-4
solution and so on; our results suggest that  in that case a cascade
of period doubling supercritical  bifurcations emerges until, at
some  $\nu$, there is a transition to global chaos. The period two
solution also supports a bimodal distribution, but now the weight of
the two peaks are not equal so the overall population takes
different value as the peaks switch positions. A similar scenario
happens in the period 4 regime.

Comparing our results with the deterministic system discussed in
\cite{mzm}, in the presence of strong demographic noise one may
observe the fully synchronized phase (when $\nu$ approaches one and
the map shows global chaos) but what about the turbulent phase? This phase appears in the
deterministic system when the coupling is too small and different
patches oscillate incoherently. Clearly the extinction phase is a consequence of turbulence.  However, the
transition from the turbulent phase to the cluster phase in the deterministic system occurs at much higher $\nu$ than the extinction transition
for the stochastic system (which is exponentially small in $N$).  The answer to this puzzle lies in the fact that there is no clear distinction between the cluster phase and the turbulent phase in the stochastic system.  As $N$ or $\nu$ are reduced, the distribution broadens and the identifiable peaks wash out.  Furthermore the total signal in both phases is time independent.
Only in the infinite $N$ limit can a sharp distinction between the two phases be made.  In this limit, the distribution function becomes two delta function peaks at infinite $N$ in the cluster phase, and have finite support in the turbulent phase..

Let us now focus our attention on the clustering phase of the deterministic system..
 One of the features of
this phase is observed here: the system clusters spontaneously into
two peaks and the global population follows a periodic orbit.
However, there are two important differences: first, in the
deterministic system for the same $\nu$ many possible solutions
exist, each corresponds to different height ratio between the peaks,
while here for any migration parameter only one stable  solution
survives, so the infinite degeneracy that characterizes the
deterministic dynamics disappears. Second,  the sharp (delta) peaks
of the deterministic solutions are replaced by smooth distributions.
One expects, however, that the stochastic system converges to the
deterministic one as $N$ increases. How does that happen?

Let us start to increase $N$.  For $N=10$ (results not shown) the
bifurcation from period 1 to period 2 is backward (subcritical),
while the bifurcation from period 2 to period 4 is still forward
even at $N=20$. The location of the bifurcation from period 1 to 2
is almost independent of $N$,  but the bifurcation from 2 to 4 is
strongly $N$ dependent, moving to smaller $\nu$ as $N$ increases. In
fact, by $N=40$ it has already moved to the backward branch of the
period 2 solution.  This situation is summarized in the upper panel
of Fig. 2, where the period 1, 2 and 4 solution branches are traced
out for $N=60$.

\begin{figure}\label{fig60}
\vspace*{-.4in}\includegraphics[width=0.47\textwidth]{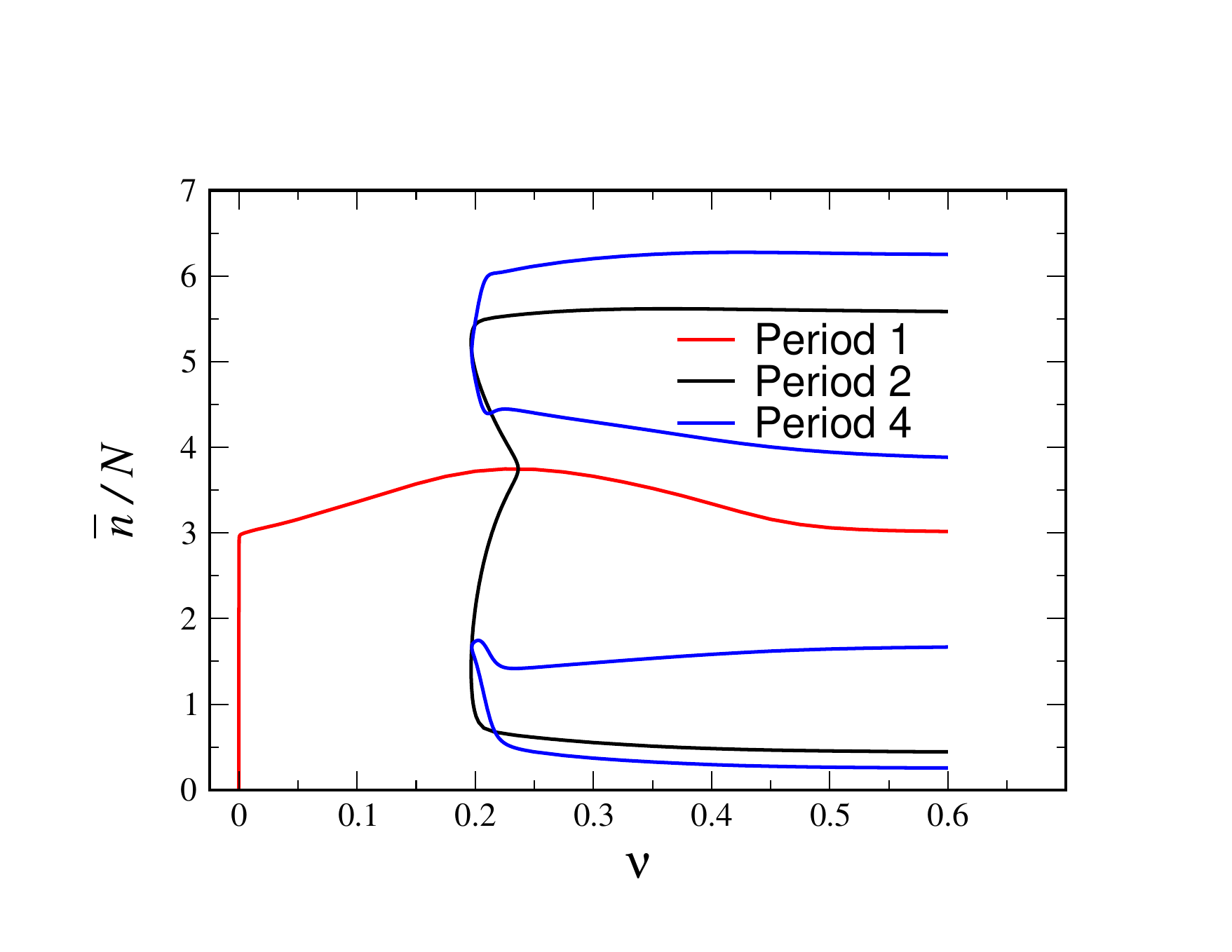}\\
\vspace*{-.5in}\includegraphics[width=0.47\textwidth]{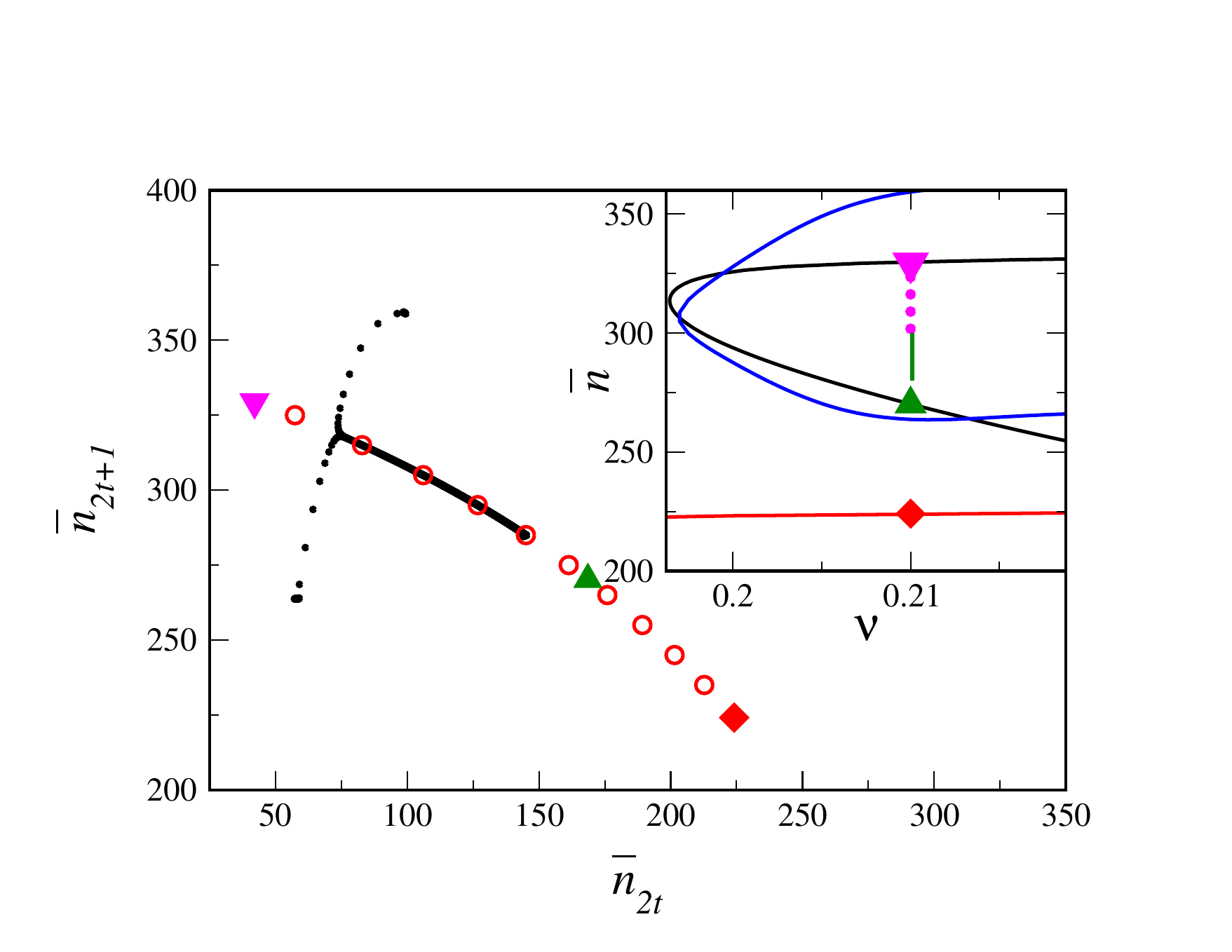}\vspace*{-.2in}
\caption{Upper:  Solution branches for $N=60$, showing the period 1,
2 and 4 branches as a function of $\nu$.  Lower: The return map
$\overline{n}_{2t+1}$ versus $\overline{n}_{2t}$, indicated by black
dots, for $N=60$, $\nu=0.21$, taken from the simulation depicted in
Fig. \ref{slowtime}.  The trajectory - a drift away from the green
triangle, then split to  the period 4 orbit - is determined by the
line of quasi solutions indicated by the red circles.  The triangles
represent true period 2 solutions for which $\alpha = 0$, and the
period 1 solution is indicated by a red diamond.  The inset is a
blowup of the relevant section of the upper panel, indicating the
true solutions and the slow flow through the quasi-solutions. The
solutions for $\nu=0.21$ are marked by the same symbols as in the
main figure. For these parameters the dynamics close to the green
triangle is so slow that measuring the dynamics becomes impractical.
When $N \to \infty$, all the region between the green triangle and
the splitting point (indicated by full green line in the inset)
becomes marginally stable.  } \label{return}
\end{figure}

The most important  issue is how the deterministic continuum of
period two solutions is recovered as $N$ grows to infinity.  As
explained above the period 2 solutions are identified by searching
for all pairs of $\lambda_1,\lambda_2$ that admit an invariant
eigenvector $\psi$  such that
 $\lambda_1 = \sum \psi_n F(n)$,
and $\lambda_2 = \sum F(n) \left[{\cal{M}}(\lambda_1)
\psi\right]_n$. It turns out  that, for large $N$, there is a range
of $\lambda_1$, $\lambda_2$ for which the Markov matrix
${\cal{M}}_{21} \equiv {\cal{M}}(\lambda_2){\cal{M}}(\lambda_1)$
admits, in addition to its invariant eigenvector, an additional
eigenstate $\tilde{\psi}$ with an eigenvalue very close to 1, say,
$1-\epsilon$. Thus, up to a small term (for $N=80$, e.g., $\epsilon
= 10^{-12}$) any linear combination of the first and the second
eigenvectors imitates the real invariant state until $t \sim
1/\epsilon$. Within this time horizon one has, effectively, a
\emph{continuous family} of invariant eigenvectors of
${\cal{M}}_{21}$, $(1-\alpha) \psi + \alpha \tilde{\psi}$. The two
auxiliary conditions no longer are sufficient to determine a
solution.  We call these solutions for which we ignore $\epsilon$ a
\emph{quasi-solution}, of which there exists a continuous family
depending on $\alpha$. It turns out that $\epsilon$ decreases
sharply with increasing $N$; the deterministic limit emerges from
this continuous family of solutions as explained below.

\begin{figure}
\includegraphics[width=0.47\textwidth]{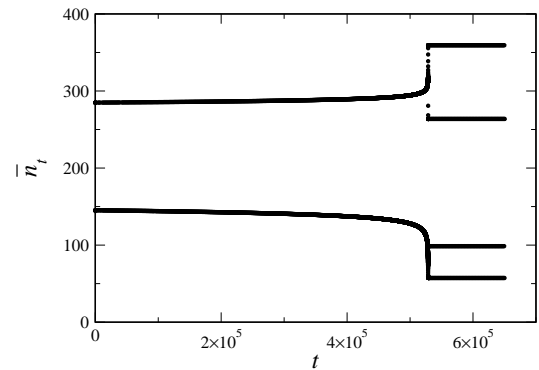}
\caption{The average population $\overline{n}_t$ versus $t$ for
$N=60$, $\nu=0.21$. Data obtained by direct integration of the
Master equation.} \label{slowtime}
\end{figure}

In Fig. \ref{slowtime}, we see all this exemplified in a simulation,
where we plotted $\overline{n}_t$ as a function of $t$, for $N=60$,
$\nu=0.21$.  We see that for times less than roughly $5\cdot 10^5$,
the system exhibits an essentially period 2 type behavior, with an
extremely slow drift of the two states.  Suddenly, beyond this
point, the system converts to a period 4 behavior.  A good way of
analyzing the drift is to plot $\overline{n}_{2t+1}$ vs.
$\overline{n}_{2t}$, as seen in Fig. \ref{return} (lower panel).  If
the system had a true period 2 orbit, this graph would show a single
point. Instead, the drift converts this into a curve.  The points on
this curve coincide precisely with the above described
quasi-solutions, a number of which are indicated by circles.  The
true solutions are represented by triangles (period 2) and a diamond (period 1) in the figure.  The system
drifts to larger amplitude oscillations, until the instability is
encountered and the it goes to a period 4 orbit, represented by two
dots in the figure.  While there are quasi-solutions (as well as a
true solution $\alpha = 0$) beyond this point, they are not
dynamically relevant due to their strong instability.

In summary, then, we have seen that adding demographic noise to a
globally coupled chaotic map has a marked effect on the dynamics,
leading in fact to very regular dynamics for intermediate coupling
strength.  As the noise strength is reduced, there appears an
exponentially long scale, which goes over to the continuous family
of solutions seen in the no noise limit.

Beyond the problems considered here, our work suggests a novel
mechanism for stochastic-deterministic ("quantum-classical")
correspondence. In general it is assumed that the infinite number of
solutions that characterizes the deterministic case emerge  from the
discrete eigenstates of the  stochastic theory when  the level
spacing approaches zero in the weak stochasticity limit. Here we
observed a different scenario, where only two eigenstates provide us
with a continuum of deterministic solutions at the $N \to \infty$
limit.

\end{document}